\documentclass[number,citesort,dvips]{arxbj}
\usepackage{subeqn}
\usepackage{colortbl}
\usepackage{graphicx}
\aid{0}
\volume{17}
\issue{3}
\pubyear{2011}
\firstpage{827}
\lastpage{844}
\doi{10.3150/10-BEJ300}

\makeatletter
\newtheorem{theo}{Theorem}

\newtheorem{cor}{Corollary}
\newtheorem{lemma}{Lemma}

\newproclaim{defn}{Definition}
\newremark{exa}{Example}
\newremark{rem}{Remark}

\newcommand{\eqref}[1]{(\ref{#1})}
\newcommand\Perp{\protect\mathpalette{\protect\independenT}{\perp}}
\def\independenT#1#2{\mathrel{\rlap{$#1#2$}\mkern4.1mu{#1#2}}}

\newcommand{\ci}{\Perp}
\renewcommand{\b}[1]{\mathbf{#1}}

\newcommand{\n}[0]{\hspace*{.35em}}
\newcommand{\nn}[0]{\hspace*{.7em}}
\newcommand{\dal}{\mbox{$  \frac{\n}{\n}\frac{\; \,}{\;}  \frac{\n}{\n}$}}
\newcommand{\fra}{\mbox{$\hspace{.05em} \frac{\nn\nn}{\nn}\!\!\!\!\! \succ \,$}}

\newcommand{\ical}{\mathcal{I}}

\newcommand{\tcal}{\mathcal{T}}

\newcommand{\pa}{\operatorname{pa}}
\newcommand{\pre}{\operatorname{pre}}
\newcommand{\setdiff}{\setminus}
\newcommand{\prodx}{\prod}
\makeatother

\begin{document}
\begin{frontmatter}

\title{Chain graph models
of multivariate regression type for categorical data}
\runtitle{Chain graph models of multivariate regression type}

\begin{aug}
\author[a]{\fnms{Giovanni M.} \snm{Marchetti}\corref{}\thanksref{a}\ead[label=e1]{giovanni.marchetti@ds.unifi.it}}
\and
\author[b]{\fnms{Monia} \snm{Lupparelli}\thanksref{b}\ead[label=e2]{monia.lupparelli@unibo.it}}
\runauthor{G.M. Marchetti and M. Lupparelli}
\address[a]{Dipartimento di Statistica ``G. Parenti'', University of
Florence,
Viale Morgagni 59,
50134 Firenze, Italy. \printead{e1}}
\address[b]{Dipartimento di Scienze Statistiche ``P. Fortunati'',
University of Bologna,
Via Belle Arti 41,
40126 Bologna, Italy. \printead{e2}}
\end{aug}

\received{\smonth{6} \syear{2009}}
\revised{\smonth{4} \syear{2010}}

%
\begin{abstract}
We discuss a class of chain graph models for categorical variables
defined by what we call a multivariate regression chain graph Markov property.
First, the set of local independencies of these models is shown to be
Markov equivalent to
those of a chain graph model recently defined in the literature. Next
we provide a parametrization based on a sequence of generalized linear
models with a
multivariate logistic link function that captures all independence
constraints in any
chain graph model of this kind.
\end{abstract}

%

\end{frontmatter}

\section{Introduction}

Discrete graphical Markov models are models for discrete distributions
representable by graphs, associating nodes with the variables and
using rules that translate properties of the graph into conditional
independence statements between variables.
There are several classes of graphical models;
see \citep{wercox2004} for a review. In this paper we focus on the class
of multivariate regression chain graphs and we discuss
their definition and parametrization
for discrete variables.

Multivariate regression chain graphs generalize directed acyclic
graphs, which
model recursive sequences of univariate responses, by allowing multiple
responses. As in all chain graph models the variables can be arranged
in a sequence of
blocks, called chain components, ordered on the basis
of subject-matter considerations, and the variables
within a block are considered to be on an equal standing as responses.
The edges are undirected within the chain components, drawn
as dashed lines \citep{cw1996} or
as bi-directed arrows \citep{ric2003}, and
directed between components, all pointing in the same direction, that is,
with no chance of semi-directed cycles. One special feature of
multivariate regression chain graphs is that the responses
are potentially depending on all the variables in all previous groups,
but not on the
other responses.
Chain graphs with this interpretation
were proposed first by Cox and Wermuth in \cite{cw1993}, with several examples
in \citep{cw1996}, Chapter~5.

%
\begin{figure}

\includegraphics{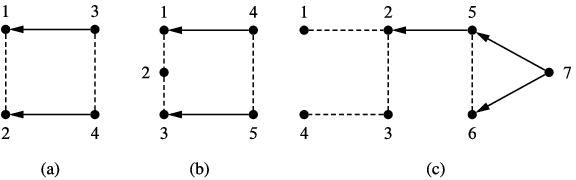}

\caption{Three chain graphs with chain components  (a)  $\tcal=\{\{
1,2\}, \{3,4\}\}$;
 (b)~$\tcal=\{\{1,2,3\},\break \{4,5\}\}$;
 (c)~$\tcal=\{\{1,2,3,4\}, \{5,6\}, \{7\}\}$.
Dashed lines only occur within chain components.}
\label{fig:cg}
\end{figure}

In the special case of a single group of responses with no
explanatory variables, multivariate regression chain graphs reduce to
covariance graphs, that is, to undirected graphs representing
marginal independencies with the basic rule that if its subgraph is
disconnected, that is,
composed by completely separated sets of nodes, then
the associated variables are
jointly independent; see \citep{drtric2008} and \citep{lupmarber2009}.
In the general case, the interpretation of the
undirected graphs within a chain component is that of a covariance graph,
but conditional on all variables in preceding components.
For example, the missing edge $(1,3)$ in the graph of Figure~\ref{fig:cg}(b)
is interpreted as the independence statement $X_1 \ci X_3 |X_{4},X_{5}$,
compactly written in terms of nodes as $1 \ci3 | 4, 5$.

The interpretation of the directed edges is that
of multivariate regression models, with a missing edge denoting a conditional
independence of the response on a variable given all the
remaining potential explanatory variables.
Thus, in the chain graph of Figure~\ref{fig:cg}(a) the missing arrow $(1,4)$
indicates the independence statement $1 \ci4 |3$.
The interpretation differs from that of classical chain graphs
(\citep{lauwer1989,fry1990}; LWF for short) where
the missing edges mean conditional independencies given all the
remaining variables,
including the other responses within the same block. However, in studies
involving longitudinal data, such as
the prospective study of
child development discussed in \citep{laucht2009}, where there are
blocks of
joint responses recorded at ages of three months, two years and four years,
an analysis conditioning exclusively on the previous
developmental states is typically appropriate.

Recently, \citep{drton2009}
distinguished four types of chain graphs comprising
the classical and the alternative \citep{amp2001}
chain graph models, called type I and II, respectively.
In this paper we give a formal definition of multivariate regression
chain graph models
and we prove that they are equivalent to the chain graph models of type
IV, in Drton's
classification \citep{drton2009}. Moreover, we provide a
parametrization based on recursive
multivariate logistic regression models. These models, introduced in
\citep{mccnel1989}, Section 6.5.4, and \citep{glomcc1995} can be
used to define all the independence constraints.
The models can be defined by an
intuitive rule, see Theorem~\ref{prop:base}, based on the structure
of the chain graph, that can be translated into a~sequence of
explicit regression models. One
consequence of the given results is that
any discrete multivariate regression chain graph model is a curved
exponential family, a~result obtained in \cite{drton2009} with a
different proof.

\section{Multivariate regression chain graphs}
The basic definitions and notation used in this paper closely follow
\cite{drton2009},
and they are briefly recalled below.
A chain graph $G = (V , E)$ is a graph with finite node set $V = \{1,
\ldots , d\}$
and an edge set $E$ that may contain
either directed edges or undirected edges. The graph has no
semi-directed cycle, that is, no
path from a node to itself with at least one directed edge such that
all directed edges have the same
orientation.
The node set $V$ of a chain graph
can be partitioned into disjoint subsets $T\in\tcal$ called \textit
{chain components},
such that all edges in each subgraph $G_T$ are undirected and the edges between
different subsets $T_1 \ne T_2$ are directed, pointing in the same direction.
For chain graphs with the multivariate regression interpretation, the
subgraphs $G_T$ within each chain component
have undirected dashed ($\dal$) or bi-directed ($\longleftrightarrow$) edges.
The former convention is adopted in this paper. Thus, the chain graph of
Figure~\ref{fig:cg}(c) has three chain components, while the previous
ones have two components.

Given a subset $A\subseteq T$ of nodes within a chain component, the
subgraph $G_A$
is said to be \textit{disconnected} if there exist two nodes in $A$
such that no path in $G_A$ has those nodes as endpoints.
In this case, $A$ can be partitioned uniquely into a set of $r > 1$
connected components $A_1, \ldots , A_r$. Otherwise, the subgraph $G_A$
is \textit{connected}.
For example,
in chain graph (c) of Figure~\ref{fig:cg}, the subgraph $G_A$ with $A =
\{1,2,4\}$ is disconnected
with two connected
components $A_1 = \{1,2\}$ and $A_2 = \{4\}$. On the other hand, the
subgraph $G_A$ with $A= \{1,2,3\}$
is connected. In the remainder of the paper, we shall say for short
that a subset $A$ of nodes in a component is
connected (respectively, disconnected) if the subgraph $G_A$ is
connected (respectively, disconnected).

\begin{figure}[t]

\includegraphics{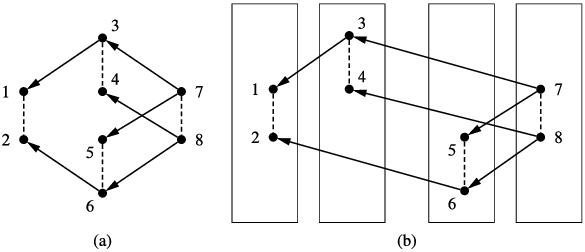}

\caption{{{(a)}} A chain graph and
 {{(b)}} one possible consistent ordering of the four chain components:
$\{1,2\} \prec\{3,4\} \prec\{5,6\} \prec\{7,8\} $.
In (b) the set of predecessors of $T = \{1,2\}$ is
$\pre(T) = \{3,4,5,6,7,8\}$, while the set of parent components of $T$
is $\pa_D(T) = \{3,4,5,6\}$. }
\label{fig:ord}
\end{figure}

Any chain graph yields a directed acyclic graph $D$ of its
chain components having $\tcal$ as a node set
and an edge $T_1 \fra T_2$ whenever there exists in the chain graph $G$
at least one edge
$v \fra w$ connecting a node $v$ in $T_1$ with a node $w$ in $T_2$.
In this directed graph, we may define for each $T$ the set $\pa_D(T)$
as the
union of all the chain components that are parents of $T$ in the
directed graph $D$.
This concept is distinct from the usual notion of the \textit{parents}
$\pa_G(A)$
of a set of nodes $A$ in the chain graph, that is, the set of all the
nodes $w$
outside $A$ such that $w \fra v$ with $v \in A$.
For instance, in the graph of Figure~\ref{fig:ord}(a), for $T = \{1,2\}$,
the set of parent components is $\pa_D(T) = \{3,4,5,6\}$, whereas the
set of parents
of $T$ is $\pa_G(T) = \{3,6\}$.

%

In this paper we start the analysis from a given chain graph $G = (V, E)$
with an associated collection $\tcal$ of chain components.
However, in applied work, where variables are linked to nodes by the
correspondence $X_v$ for $v \in V$,
usually a set of chain components is assumed known
from previous studies of substantive theories or from
the temporal ordering of the variables. For variables within such
chain components no direction of influence is specified and they are considered
as joint responses, that is, to be on equal standing.
The relations between variables in different chain components
are directional and are typically based on a preliminary distinction of
responses, intermediate responses and purely explanatory factors.
Often, a full ordering of the components
is assumed based on time order or on a subject matter working hypothesis;
see \cite{cw1996}.

Given a chain graph $G$ with chain
components $(T \mid T \in\tcal)$,
we can always define a~strict total order $\prec$ of the chain components
that is \textit{consistent} with the partial order induced by the chain graph,
such that if $T \prec T'$ then $T \notin\pa_D(T')$.
For instance, in the chain graph of Figure~\ref{fig:ord}(a) there are
four chain components ordered in graph (b) as $\{1,2\} \prec\{3,4\}
\prec\{5,6\} \prec\{7,8\} $.
Note that the chosen total order of the chain components
is in general not unique and that another consistent
ordering could be $\{1,2\} \prec\{5,6\} \prec\{3,4\} \prec\{7,8\} $.

In the remainder of the paper we shall assume that a consistent
ordering $\prec$ of the
chain components is given.
Then, for each $T$, the set of
all components preceding $T$ is known and we may define the cumulative set
$
\pre(T) = \bigcup_{T \prec T'} T'
$
of nodes contained in the predecessors of component $T$ that we
sometimes also call
the past of $T$. The set
$\pre(T)$ captures the notion of all the potential explanatory
variables of
the response variables within $T$.
By definition, as the full ordering of the components is consistent
with $G$,
the set of predecessors $\pre(T)$ of each chain component $T$ always includes
the parent components $\pa_D(T)$.

The following definition explains the meaning of the multivariate regression
interpretation of a chain graph.
\begin{defn}\label{def:1}
Let $G$ be a chain graph with chain components $(T \mid T \in\tcal)$ and let
$\pre(T)$ define an ordering of the chain components consistent with
the graph.
A joint distribution P of the random vector $\b X$ obeys the (global)
multivariate regression Markov property
for $G$ if it satisfies the following independencies. For all $T \in
\tcal$ and for
all $A \subseteq T$:
\begin{enumerate}[]
\item[\hspace*{-6pt}(\textsc{mr1})]
 if $A$ is connected:
$A \ci[\pre(T) \setminus\pa_G(A)] \mid \pa_G(A)$.
\item[\hspace*{-6pt}(\textsc{mr2})]
 if $A$ is
disconnected with connected components $A_1, \ldots , A_r:
  A_1 \ci\cdots\ci A_r \mid\pre(T)$.
\end{enumerate}
\end{defn}

Assuming that the distribution $P$ has a density $p$ with respect to a
product measure,
the definition can be stated by the following two equivalent conditions:
\begin{subequations}
%
\begin{equation}
p_{A|\pre(T)} = p_{A|\pa_G(A)}  \label{eq:mr1}
\end{equation}
for all $T$ and for all connected subset $A \subseteq T$.
%
\begin{equation}
p_{A|\pre(T)} = \prod_j p_{A_j|\pre(T)}  \label{eq:mr2}
\end{equation}
\end{subequations}
\noindent for all $T$ and for all disconnected subset $A \subset T$
with connected components
$A_j,  j = 1, \ldots , r$.

In other words, for any connected subset $A$ of responses in a
component $T$, its conditional distribution
given the variables in the past
depends only on the parents of $A$. On the other hand, if $A$ is disconnected
(i.e., the subgraph $G_A$ is disconnected)
the variables in its connected components
$A_1, \ldots , A_r$, are jointly independent given the variables in the past.
\begin{rem}
Definition~\ref{def:1} gives a local Markov property that always
implies the
following pairwise Markov property: For every uncoupled pair of nodes
$i, k$,
%
\begin{equation}
i \ci k |\pre(T),  \qquad \mbox{if } i,k \in T; \qquad i \ci k|\pre(T)\setminus
\{k\},  \qquad \mbox{if } i \in T, k \in\pre(T).
\label{eq:pair}\vadjust{\goodbreak}
\end{equation}
In particular, two pairwise independencies $i \ci k|\pre(T)$ and $i \ci
\ell|\pre(T)$
can occur only in combination with the joint independence $i \ci k,
\ell|\pre(T)$. This means
that in the associated model the composition property is always
satisfied; see
\cite{studeny2005}. Thus, even though we concentrate in this paper on
the family
of multinomial distributions that does not satisfy the composition
property, the
models in which (\textsc{mr1}) and (\textsc{mr2}) hold have this property.
\end{rem}

\begin{rem}
One immediate consequence of Definition~\ref{def:1}
is that if the
probability density $p(\b x)$ is strictly positive, then
it factorizes according to the directed acyclic graph of the chain components:
%
\begin{equation}
p(\b x) = \prod_{T \in\tcal} p\bigl(\b x_T | \b x_{\pa_D(T)}\bigr). \label{eq:fact0}
\end{equation}
This factorization property is shared by all types of chain graphs; see
\citep{wercox2004} and \citep{drton2009}.
\end{rem}

Recently, \citep{drton2009} discussed four different block-recursive
Markov properties for chain graphs, of which we discuss here those with
the \textit{Markov property of type IV}.
To state it,
we need two further concepts from graph theory. Given a chain graph $G$,
the set $\operatorname{Nb}_G(A)$
is the union of $A$ itself and the set of nodes $w$ that are \textit
{neighbours} of $A$, that is,
coupled by an undirected edge
to some node $v$ in $A$. Moreover, the set of \textit{non-descendants}
$\operatorname{nd}_D(T)$ of
a chain component $T$, is the union of all components $T'$ such that
there is no
directed path from $T$ to $T'$ in the directed graph of chain
components $D$.
%
\begin{defn}[(Chain graph Markov property of type IV  \citep
{drton2009})] \label{def:drton}
Let $G$ be a chain graph with chain components $(T\mid T \in\tcal)$
and directed acyclic graph $D$ of components. The joint probability
distribution of
$\b X$ obeys the block-recursive Markov property of type IV if it
satisfies the following independencies:
\begin{enumerate}[(\textsc{iv2})]
\item[(\textsc{iv0})]   $A \ci[\operatorname{nd}_D(T) \setminus\pa_D(T)]
\mid\pa_D(T) \mbox{ for all } T \in\tcal$;
\item[(\textsc{iv1})]   $A \ci[\pa_D(T) \setminus\pa_G(A)] \mid\pa_G(A)
\mbox{ for all } T \in\tcal\mbox{ for all } A \subseteq T$;
\item[(\textsc{iv2})]    $A \ci[T \setminus\operatorname{Nb}_G(A)] \mid
\pa_D(T) \mbox{ for all } T \in\tcal\mbox{ for all connected subsets
} A \subseteq T.$
\end{enumerate}
\end{defn}

Then we have the following result, proved in the \hyperref[appm]{Appendix}.
\begin{theo} \label{theo:1}
Given a chain graph $G$, the multivariate regression Markov
property is equivalent to the block-recursive
Markov property of type IV.
\end{theo}

This result shows that the
block-recursive property of a chain graph of type IV is in fact
simplified by Definition~\ref{def:1}.
On the other hand, Definition~\ref{def:1} depends only apparently on
the chosen full ordering of the chain components, because the
equivalent Definition~\ref{def:drton} depends only on the underlying
chain graph $G$.
\begin{exa}
The independencies implied
by the multivariate regression chain graph Markov property are
illustrated below
for each of the graphs of Figure~\ref{fig:cg}.

Graph (a) represents the independencies of the seemingly unrelated
regression model; see \citep{cw1993} and \cite{drtrich2004}.
For $T = \{1,2\}$ and $\pre(T) = \{3,4\}$ we have the independencies
$1 \ci 4 | 3$ and $2 \ci 3 | 4$. Note that for the connected set $A =
\{1,2\}$
the condition (\textsc{mr1}) implies the trivial statement
$A \ci\varnothing| \pre(T)$.

In graph (b) one has $T = \{1,2,3\}$ and $\pre(T) = \{4,5\}$. Thus, for
each connected
subset $A\subseteq T$, by (\textsc{mr1}), we have the non-trivial statements
\[
1 \ci5 |4; \qquad   2 \ci4, 5; \qquad
 3 \ci4 |5;  \qquad  1, 2 \ci5 |4; \qquad
 2, 3 \ci4 |5.
\]
Then, for the remaining disconnected set $A = \{1,3\}$
we obtain by (\textsc{mr2}) the independence $1 \ci3 |4, 5$.

In graph (c), considering the conditional distribution
$p_{T|\pre(T)}$ for $T = \{1,2,3,4\}$ and $\pre(T) = \{5,6,7\}$,
we can define independencies for each of the eight connected
subsets of $T$. For instance, we have
\[
1 \ci5, 6, 7;   \qquad  1, 2 \ci6, 7 |5;  \qquad  1, 2, 3, 4 \ci
7 |5, 6.
\]
The last independence is equivalent to the
factorization
$
p = p_{1234|56}\cdot p_{56|7}\cdot p_{7}
$
of the joint probability distribution according to the directed acyclic
graph of the chain components.
The remaining five disconnected subsets of $T$ imply the conditional
independencies
$1, 2 \ci4 | 5, 6, 7$ and $1 \ci3, 4 |5, 6, 7.$ Notice that
when in a component there are two uncoupled nodes, then there is a conditional
independence given simply the common parents of the two nodes. For
example, in graph (c),
we have not only $1 \ci3|5, 6$ but also $1 \ci3 |5$.
\end{exa}

\begin{rem}\label{rem:dec}
When each component $T$ induces a complete
subgraph $G_T$
and, for all subsets $A$ in $T$, the set of parents of $A$, $\pa_G(A)$,
coincides
with the set of the parent components of $T$, $\pa_D(T)$, then the
only conditional independence implied by the
multivariate regression Markov property is
\[
A \ci[\pre(T) \setminus\pa_D(T)] | \pa_D(T) \qquad  \mbox{for all } A
\subseteq T,\ T \in\tcal.
\]
This condition is in turn equivalent just to the factorization \eqref
{eq:fact0} of the
joint probability distribution.
\end{rem}

\begin{rem}
In Definition~\ref{def:1},
(\textsc{mr2}) is equivalent to imposing that
for all $T$ the conditional distribution $p_{T|\pre(T)}$
satisfies the independencies of a
covariance graph model with respect to the subgraph $G_T$.

In \citep{lupmarber2009}, Proposition 3, it is shown that
a covariance graph model is defined by constraining to zero, in the
multivariate logistic parametrization, the parameters corresponding to all
disconnected subsets of the graph.
In the following subsection we extend this approach to the multivariate
regression
chain graph models.
\end{rem}

%
\section{Recursive multivariate logistic regression models}
\subsection{Notation}
Let $\b X = (X_v\mid v \in V)$ be a discrete random vector, where each variable
$X_v$ has a finite number $r_v$ of levels. Thus $\b X$ takes values in
the set
$\ical= \prod_{v \in V} \{1, \ldots , r_v\}$ whose elements are
the cells of the joint contingency table, denoted by
$\b i = (i_1, \ldots , i_d)$. The first level of each variable is considered
a reference level and we consider also the set $\ical^\star=
\prod_{v \in V} \{2, \ldots , r_v\}$ of cells having all indices
different from the first. The elements of $\ical^\star$ are denoted by
$\b i^\star$.

The joint probability distribution of $\b X$ is defined by the mass function
\[
p(\b i) = P(X_v = i_v, v = 1, \ldots , d) \qquad  \mbox{for all } \b i \in\ical,
\]
or equivalently by the probability vector $\b p = (p(\b i), \b i \in
\ical)$.
With three variables we shall use often $p_{ijk}$ instead of $p(i_1,i_2,i_3)$.

Given two disjoint subsets $A$ and $B$ of $V$,
the marginal probability distribution of $\b X_B$ is
$
p(\b i_B) = \sum_{\b j_B = \b i_B} p(\b j)
$
where $\b i_B$ is a subvector of $\b i$ belonging\vspace*{1pt} to the marginal
contingency table
$\ical_B = \prod_{v \in B} \{1, \ldots , r_v\}$.
The conditional probability
distributions are defined as usual and denoted by
$p(\b i_A|\b i_{B})$, for $\b i_A \in\ical_A$ and $\b i_B \in\ical_B$
or, compactly, by $p_{A|B}$. When appropriate, we define
the set $\ical_B^\star= \prod_{v \in B} \{2, \ldots , r_v\}$.

A discrete multivariate regression chain graph model $\mathbf
{P}_{\mathrm{MR}}(G)$
associated with the chain graph
$G = (V, E)$ is the set of strictly positive joint probability
distributions $p(\b i)$ for $i \in\ical$
that obeys the multivariate regression Markov property. By Theorem~\ref
{theo:1} this class
coincides with the set $\mathbf{P}_{\mathrm{IV}}(G)$ of discrete chain
graph models of type IV.

In the next subsection we define an appropriate parametrization for
each component of the standard
factorization
%
\begin{equation} \label{eq:fact1}
p(\b i) = \prod_{T \in\tcal} p\bigl(\b i_T |\b i_{\pre(T)}\bigr)
\end{equation}
of the joint probability distribution. Actually we define a saturated
linear model for a~suitable transformation of the parameters of each
conditional probability distribution $p(\b i_T |\b i_{\pre(T)})$.

\subsection{Multivariate logistic contrasts}
The suggested link function is the multivariate logistic
transformation; see
\citep{mccnel1989}, page~219, and \citep{glomcc1995}. This
link transforms the joint probability vector of the responses into
a vector of logistic contrasts defined for all the marginal distributions.
The contrasts of interest are all sets of univariate, bivariate and
higher order
contrasts. In general, a \textit{multivariate logistic contrast} for a
marginal table
$\b p_A$ is defined by
the function
%
\begin{equation}
\eta^{(A)}(\b i_A^\star) = \sum_{s \subseteq A} (-1)^{|A\setdiff s|}
\log p(\b i_s^\star, \b1_{A \setdiff s})  \qquad  \mbox{for } \b i^\star\in
\ical^\star_A,
\end{equation}
where the notation $|A \setminus s|$ denotes the cardinality of set $A
\setminus s$.
The contrasts for a margin $A$ are denoted by $\bolds\eta^{(A)}$ and
the full vector of the contrasts for all non-empty margins $A \subseteq
V$ are
denoted by $\bolds\eta$.
The following example illustrates the transformation for two responses.

\begin{exa} \label{ex:mlogit}
Let $p_{ij}$, for $i = 1,2$, $j = 1,2,3$ be a joint bivariate distribution
for two discrete variables $X_1$ and $X_2$.
Then the multivariate logistic transform changes the
vector $\b p$ of probabilities, belonging to the 5-dimensional simplex,
into the $5 \times1$ vector
\[
\bolds\eta= \pmatrix{
\bolds\eta^{(1)}\cr \bolds\eta^{(2)}\cr \bolds\eta^{(12)}}, \qquad
 \mbox{where }
\bolds\eta^{(1)} = \log\frac{p_{2+}}{p_{1+}},
\bolds\eta^{(2)} = \pmatrix{\displaystyle
\log\frac{p_{+2}}{p_{+1}}\cr\displaystyle
\log\frac{p_{+3}}{p_{+1}}
},
\bolds\eta^{(12)} = \pmatrix{\displaystyle
\log\frac{p_{11}p_{22}}{p_{21}p_{12}}\cr\displaystyle
\log\frac{p_{11}p_{23}}{p_{21}p_{13}}
},
\]
where the $+$ suffix indicates summing over the corresponding index.
Thus, the parameters $\bolds\eta^{(1)}$ and $\bolds\eta^{(2)}$ are
marginal baseline logits
for the variables $X_1$ and $X_2$, while $\bolds\eta^{(12)}$ is a~%
vector of
log odds ratios. The definition used in this paper uses baseline coding,
that is, the contrasts are defined with respect to a reference level, by
convention the first.
Therefore the dimension of
the vectors $\bolds\eta^{(1)}$, $\bolds\eta^{(2)}$ and $\bolds\eta^{(12)}$ are
the dimensions of the sets $\ical^\star_1$, $\ical^\star_2$ and $\ical
^\star_{12}$.
Other coding schemes can be adopted, as discussed, for instance, in
\citep{wercox1992}
and \citep{barcolfor2007}.
\end{exa}

\begin{rem}\label{rem:diff}
This transformation for multivariate binary
variables is discussed in \citep{glomcc1995}, where it is shown that
the function from $\b p$ to $\bolds\eta$ is a smooth
($C^\infty$) one-to-one function having a
smooth inverse, that is, it is a diffeomorphism; see also \citep{berrud2002}.
For general discrete variables, see \citep{lupmarber2009}.
The parameters are not variation-independent, that is, they do not
belong to a hyper-rectangle.
However, they satisfy the
\textit{upward compatibility property},
that is, they have the same meaning across different marginal distributions;
see \citep{glomcc1995} and \citep{lupmarber2009}, Proposition 4.
Often the multivariate
logistic link is written as
%
\begin{equation}
\bolds\eta= \b C \log(\b M \b p), \label{eq:CM}
\end{equation}
where $\b C$ and $\b M$ are suitable Kronecker products of contrast and
marginalization matrices,
respectively. For the explicit construction of these matrices, see \cite
{barcolfor2007}.
\end{rem}

\subsection{Saturated model}
We specify the dependence of the responses in each component $T$
on the variables in the past by defining a saturated multivariate
logistic model for
the conditional probability distribution $p_{T|\pre(T)}$.
The full saturated model for the joint probability $p$
then follows from the factorization~\eqref{eq:fact1}.

For each covariate class $\b i_{\pre(T)} \in\ical_{\pre(T)}$,
let $\b p(\b i_{\pre(T)})$ be the vector with strictly positive
components $p(\b i_T | \b i_{\pre(T)})> 0 $ for $\b i_T \in\ical_T$.
Then consider the associated
conditional multivariate logistic parameters
$\bolds\eta(\b i_{\pre(T)})$ defined using the link function \eqref
{eq:CM}. Notice that
this vector is composed of contrasts $\bolds\eta^{(A)}(\b i_{\pre(T)})$
for all non-empty subsets $A$ of $T$.
Then we express the dependence of each of them on
the variables in the preceding components
by a complete factorial model
%
\begin{equation}
\bolds\eta^{(A)}\bigl(\b i_{\pre(T)}\bigr) = \sum_{b \subseteq\pre
(T)} \bolds\beta^{(A)}_b(\b i_b).
\label{eq:satmo}
\end{equation}
Here the vectors $\bolds\beta^{(A)}_b(\b i_b)$ have dimensions of the sets
$\ical_A^\star$, and are defined according to the baseline coding, and
thus vanish
when at least one component of $\b i_b$ takes on the first level.
Again, here different codings may be used if appropriate.
Often it is useful to express \eqref{eq:satmo} in matrix form
%
\begin{equation}
\bolds\eta^{(A)} = \b Z^{(A)} \bolds\beta^{(A)}, \label{eq:eta-beta}
\end{equation}
where $\bolds\eta^{(A)}$ is the column vector obtained by stacking all
vectors $\bolds\eta^{(A)}(\b i_{\pre(T)})$ for $\b i_{\pre(T)} \in\ical
_{\pre(T)}$,
$\b Z^{(A)}$ is a full-rank design matrix and $\bolds\beta^{(A)}$
is a parameter vector.
\begin{exa}
Suppose that in Example~\ref{ex:mlogit} the responses $X_1$ and $X_2$
depend on two binary explanatory variables $X_3$ and $X_4$, with levels indexed
by $k$ and $\ell$, respectively. Then the saturated model is
\[
\bolds\eta^{(A)}(k,\ell) = \bolds\beta^{(A)}_\varnothing+ \bolds\beta
^{(A)}_3(k) + \bolds\beta^{(A)}_4(\ell) +
\bolds\beta^{(A)}_{34}(k,\ell),  \qquad   k, \ell= 1,2,
\]
for $A = \{1\}, \{2\}, \{12\}$.
The explicit form of the matrix $\b Z^{(A)}$
in equation \eqref{eq:eta-beta}
is, using the Kronecker product $\otimes$ operator,
\[
\b Z^{(A)} = \b I \otimes\pmatrix{ 1 & 0 \cr 1 & 1 }
\otimes\pmatrix{ 1 & 0 \cr 1 & 1 },
\]
that is, a matrix of a complete factorial design matrix, where $\b I$
is an identity
matrix of an order equal to the common dimension of each $\bolds\eta
^{(A)}(k,\ell)$.
Following \citep{mccnel1989}, page~222, we shall denote the model, for
the sake of brevity, by
a multivariate model formula
\[
X_1\dvtx   X_3*X_4;  \qquad   X_2\dvtx   X_3*X_4; \qquad
X_{12}\dvtx   X_3*X_4,
\]
where $X_3*X_4 = X_3 + X_4 + X_3\cdot X_4$ is the factorial expansion in
Wilkinson and Rogers' symbolic notation \citep{wilrog1973}.
\end{exa}

When we need to express the overall 1--1 smooth transformation of the
conditional probability vectors $\b p(\b i_{\pre(T)})$, denoted collectively
by $\b p_T$, into the logistic and regression parameters we
introduce the vectors $\bolds\eta_{T}$ and $\bolds\beta_T$ obtained by
concatenating the
parameters $\bolds\eta^{(A)}$ and $\bolds\beta^{(A)}$, respectively,
for all non-empty subsets $A$ of $T$, writing
%
\begin{equation}\label{eq:modT}
\bolds\eta_{T} = \b Z_T \bolds\beta_T,
\end{equation}
where $\b Z_T = \operatorname{diag}(\b Z^{(A)})$
is a full rank block-diagonal matrix of the saturated model, and
%
\begin{equation} \label{eq:fullink}
\b C_T \log(\b M_T \b p_T) = \bolds\eta_T,
\end{equation}
where $\b C_T$ and $\b M_T$ are suitable overall contrast and marginalization
matrices.

\section{Discrete multivariate regression chain graph models}
\subsection{Linear constraints}
A multivariate regression chain graph model is specified by
zero constraints on the parameters $\bolds\beta_T$ of the saturated model
\eqref{eq:modT}. We give first an example and then we state the general result.

\begin{exa} Continuing the previous example
for the chain graph $G$ of Figure~\ref{fig:cg}(a),
we shall require
that $X_1$ depends only on $X_3$ and
$X_2$ depends only on $X_4$. Therefore, we specify the
model
\begin{eqnarray*}
\bolds\eta^{(1)}(k,\ell) &=& \bolds\beta^{(1)}_\varnothing+ \bolds\beta
^{(1)}_3 (k),\\
\bolds\eta^{(2)}(k,\ell) &=& \bolds\beta^{(2)}_\varnothing+ \bolds\beta
^{(2)}_4 (\ell),\\
\bolds\eta^{(12)}(k,\ell) &=& \bolds\beta^{(12)}_\varnothing+ \bolds\beta
^{(12)}_3(k) +
\bolds\beta^{(12)}_4(\ell) +\bolds\beta^{(12)}_{34}(k,\ell)
\end{eqnarray*}
with a corresponding multivariate model formula
\[
X_1\dvtx   X_3,  \qquad   X_2\dvtx   X_4,  \qquad   X_{12}\dvtx   X_3*X_4.
\]
The reduced model satisfies the two independencies $1 \ci4 | 3$ and
$2 \ci3|4$ because the first two equations are equivalent to
$p_{1|34} = p_{1|3}$ and $p_{2|34} = p_{1|4}$, respectively.
The log odds-ratio between $X_1$ and $X_2$, on the other hand,
depends in general on all the combinations $(k, \ell)$ of levels of the
two explanatory variables.
\end{exa}

The following theorem, proved in the \hyperref[appm]{Appendix}, states a general
rule to parametrize any discrete chain graph model of the multivariate
regression type.
\begin{theo}
\label{prop:base}
Let $G$ be a chain graph and let $\pre(T)$
be a consistent ordering of the chain components $T \in\tcal$.
A joint distribution of the discrete random vector $\b X$ belongs to
$\mathbf{P}_{\mathrm{MR}}(G)$
if and only if, in the multivariate logistic model \eqref{eq:satmo},
the parameters $\bolds\beta^{(A)}_b(\b i_b) =\b0$, $\b i_b \in\ical_b$,
whenever
\begin{subequations}
%
\begin{eqnarray}
A \mbox{ is connected and }   b &\subseteq&\pre(T) \setminus\pa_G(A),
\label{eq:constr1}\\
A \mbox{ is disconnected and }   b &\subseteq&\pre(T)
\label{eq:constr2}
\end{eqnarray}
for all $A \subseteq T$ and for all $T \in\tcal$.
\end{subequations}
\end{theo}

Notice that equations \eqref{eq:constr1} and \eqref{eq:constr2}
correspond to
conditions \eqref{eq:mr1} and \eqref{eq:mr2}, respectively, of
Definition~\ref{def:1}.
Thus the multivariate regression chain graph model turns out to be
$\bolds\eta^{(A)}(\b i_{\pre(T)}) = \sum_{b \subseteq\pa_G(A)} \bolds
\beta^{(A)}_b(\b i_b)$
if $A$ is connected and $\b0$ if $A$ is disconnected.\vspace*{1pt} In matrix form
we have a linear
predictor
%
\begin{equation}
\bolds\eta_T = \b Z_r \bolds\beta_r,
\label{eq:redmod}
\end{equation}
where $\b Z_r$ is the matrix of the reduced model obtained by removing
selected columns
of $\b Z_T$, and $\bolds\beta_r$ are the associated parameters.

The proof of Theorem~\ref{prop:base} is based on a basic property of the
regression parameters
$\bolds\beta^{(A)}_b(\b i_b)$ of model \eqref{eq:satmo}, that is, that
they are
identical to log-linear parameters defined in selected marginal tables.
Specifically, each $\bolds\beta^{(A)}_b(\b i_b)$ coincides with the vector
of log-linear parameters $\bolds\lambda^{AB}_{Ab}$ of order $A \cup b$
in the marginal table $A \cup\pre(T)$.
See Lemma~\ref{prop:loglin} in the
\hyperref[appm]{Appendix}.

Theorem~\ref{prop:base} shows also that the chain graph model $\mathbf
{P}_{\mathrm{MR}}(G)$ is
defined by a set of linear restrictions on a multivariate logistic
parametrization and thus
is a curved exponential family.

\begin{exa}\label{ex:3}
From Theorem~\ref{prop:base}, the chain graph model of Figure~\ref
{fig:cg}(b) is defined by the equations
\begin{eqnarray*}
\eta^{(1)}(k,l) &=& \beta_{\phi}^{(1)} + \beta_{4}^{(1)}(k),  \qquad
\eta^{(2)}(k,l) = \beta_{\phi}^{(2)},  \qquad
\eta^{(3)}(k,l) = \beta_{\phi}^{(3)} + \beta_{5}^{(3)}(l),
\\
\eta^{(12)}(k,l) &=&\beta_{\phi}^{(12)} + \beta_{4}^{(12)}(k),  \qquad
\eta^{(13)}(k,l) = 0,  \qquad
\eta^{(23)}(k,l) = \beta_{\phi}^{(23)} + \beta_{5}^{(23)}(l),
\\
 \eta^{(123)}(k,l) &=& \beta_{\phi}^{(123)} + \beta
_{4}^{(123)}(k) + \beta_{5}^{(123)}(l) +\beta_{45}^{(123)}(k,l)
\end{eqnarray*}
and by
the multivariate logistic model formula
\[
X_1\dvtx  X_4, \qquad
X_2\dvtx  1, \qquad
X_3\dvtx  X_5, \qquad
X_{12}\dvtx  X_4 , \qquad
X_{13}\dvtx  0, \qquad
X_{23}\dvtx  X_5, \qquad
X_{123}\dvtx  X_4*X_5.
\]
Notice that the marginal logit of $X_2$ does not depend on the
variables $X_4, X_5$. This is denoted by $X_2\dvtx   1 $. On the other hand,
the missing edge $(1,3)$ with associated independence $1 \ci3|4, 5$
implies that
the bivariate logit between $X_1$ and $X_3$ is zero, denoted
by model formula $X_{13}\dvtx   0$. The above equations reflect
exactly the independence structure encoded by the multivariate
regression Markov property but leave
a complete free model for the three-variable logistic parameter $\eta^{(123)}$.

%
%
\begin{table*}[t]
\tabcolsep=0pt
\tablewidth=310pt
\caption{Marginal log-linear parameters of the saturated model
for a discrete multivariate logistic model
with three responses and two explanatory variables.
Each row lists log-linear parameters defined within a marginal
table indicated in the last column. The non-zero terms of the chain
graph model of Example~\protect\ref{ex:3}
are shown in boldface.
The shaded part of the table collects the interactions of an order
higher than two}
\label{tab:1}
\begin{tabular*}{310pt}{@{\extracolsep{\fill}}l
>{\columncolor[gray]{0.8}}l
>{\columncolor[gray]{0.8}}l
>{\columncolor[gray]{0.8}}l
>{\columncolor[gray]{0.8}}l
>{\columncolor[gray]{0.8}}l@{}}
\hline
\multicolumn{1}{>{\columncolor{white}}l}{Logit} & \multicolumn{4}{>{\columncolor{white}}l}{Parameters}
& \multicolumn{1}{>{\columncolor{white}}l@{}}{Margin}\\
\ccline{2-5}\vspace*{-8.5pt}\\
 & \multicolumn{1}{>{\columncolor{white}}l}{Const.}
 & \multicolumn{1}{>{\columncolor{white}}l}{4}
 & \multicolumn{1}{>{\columncolor{white}}l}{5}
 & \multicolumn{1}{>{\columncolor{white}}l}{45}   \\
 \hline
\multicolumn{1}{>{\columncolor{white}}l}{\hphantom{12}1} & \multicolumn{1}{>{\columncolor{white}}l}{\hphantom{12}\textbf{1}} & \multicolumn{1}{>{\columncolor{white}}l}{\hphantom{12}\textbf{14}} & \multicolumn{1}{>{\columncolor{white}}l}{\hphantom{12}15} &  \hphantom{12}145 & {\hphantom{12}145} \\
\multicolumn{1}{>{\columncolor{white}}l}{\hphantom{12}2} & \multicolumn{1}{>{\columncolor{white}}l}{\hphantom{12}\textbf{2}} & \multicolumn{1}{>{\columncolor{white}}l}{\hphantom{12}24} & \multicolumn{1}{>{\columncolor{white}}l}{\hphantom{12}25} & \hphantom{12}245 & {\hphantom{12}245} \\
\multicolumn{1}{>{\columncolor{white}}l}{\hphantom{12}3} & \multicolumn{1}{>{\columncolor{white}}l}{\hphantom{12}\textbf{3}}
& \multicolumn{1}{>{\columncolor{white}}l}{\hphantom{12}34} & \multicolumn{1}{>{\columncolor{white}}l}{\textbf{\hphantom{12}35}} & \hphantom{12}345 & {\hphantom{12}345} \\
\multicolumn{1}{>{\columncolor{white}}l}{\hphantom{1}12} & \multicolumn{1}{>{\columncolor{white}}l}{\hphantom{1}\textbf{12}} & \hphantom{1}\textbf{124} & \hphantom{1}125 & \hphantom{1}1245 & {\hphantom{1}1245} \\
\multicolumn{1}{>{\columncolor{white}}l}{\hphantom{1}13} & \multicolumn{1}{>{\columncolor{white}}l}{\hphantom{1}13} & \hphantom{1}134 & \hphantom{1}135 & \hphantom{1}1345 & {\hphantom{1}1345} \\
\multicolumn{1}{>{\columncolor{white}}l}{\hphantom{1}23} & \multicolumn{1}{>{\columncolor{white}}l}{\hphantom{1}\textbf{23}} & \hphantom{1}234 & \hphantom{1}\textbf{235} & \hphantom{1}2345 &  {\hphantom{1}2345} \\
\multicolumn{1}{>{\columncolor{white}}l}{123} & \textbf{123} & \textbf{1234} & \textbf{1235} &\textbf{12345} &
{\textbf{12345}}
\\
\hline
\end{tabular*}
\vspace*{5pt}
\end{table*}

Table~\ref{tab:1} lists the parameters (and their log-linear interpretations)
of the saturated model. The non-vanishing parameters of the chain graph model
are in boldface. The shaded portion of the table indicates the interactions
of an order higher than two. Therefore, the chain graph model contains
seven parameters
in the shaded area that have a~more complex interpretation and that are not
strictly needed to define the independence structure.
This leads us to consider, as a starting model, a multivariate logistic
regression
model with no parameters of log-linear order higher than two and then
use a
backward selection strategy to test for the independencies. Some
adjustment of the
procedure is needed to include selected higher order interactions when needed.
Notice also that the parameters in Table~\ref{tab:1} form a marginal log-linear
parametrization in the sense of Bergsma and Rudas  \cite{berrud2002},
a result that can be proved for any discrete multivariate regression
chain model.
For an example see \citep{marlup2008}.
\end{exa}

A parallel multivariate logistic parametrization for the model $\mathbf
{P}_{\mathrm{IV}}(G)$
can be obtained from Definition~\ref{def:drton} and the associated
characterization
in terms of densities of Lemma~\ref{prop:iv} in the \hyperref[appm]{Appendix}.
In this case, using the factorization \eqref{eq:fact0},
the multivariate logistic models can be defined in
the lower-dimensional conditional distributions $p_{T|\pa_D(T)}$.
Therefore we state the following
corollary.

\begin{cor}
The joint probability distribution of the random
vector $\b X$ belongs to $\mathbf{P}_{\mathrm{IV}}(G)$ if and only if
it factorizes
according to equation \eqref{eq:fact0},
and for each conditional distribution $p(\b i_T | \b i_{\pa_D(T)})$,
for $T \in\tcal$, the multivariate logistic
parameters are
%
\begin{equation}
\bolds\eta^{(A)}\bigl(\b i_{\pa_D(T)}\bigr) =
\cases{\displaystyle
\sum_{b \subseteq\pa_G(A)} \bolds\beta^{(A)}_b(\b i_b)  &  \quad  for
all connected $A\subseteq T$,\cr\displaystyle
\b0  & \quad   for all disconnected $A\subseteq T$.
}
\end{equation}
\end{cor}

In the class of models defined in Remark~\ref{rem:dec}, corresponding
exactly to the factorization~\eqref{eq:fact0}, all the independencies
are obtained by setting
$\pa_G(A) = \pa_D(T)$ for all $A \subseteq T$ in equation \eqref{eq:constr1}.

\subsection{Likelihood inference}
The estimation of discrete multivariate regression chain models can be
carried out
by fitting separate multivariate logistic regression models to each
factor $p_{T|\pre(T)}$
of the decomposition \eqref{eq:fact1}. Specifically, given a block $T$
of responses and
the group of covariates $\pre(T),$
we consider the table of frequencies $\b Y_k$ for each covariate class
$k$, where
$k = 1, \ldots , K$ is an index numbering the cells of the marginal table
$\ical_{\pre(T)}$. Then
we assume that each $\b Y_k \sim M(n_k, \b p_k)$ is multinomial with
$\b p_k = \b p(\b i_{\pre(T)})$.
Given $K$ independent observations $(\b Y_1,n_1), \ldots , (\b Y_K, n_K)$
the vector
$\b Y = \operatorname{vec}(\b Y_1, \ldots , \b Y_K)$ has a~%
product-multinomial distribution and
the log-likelihood is
%
\begin{equation}
l(\bolds\omega) = \b y^T \bolds\omega- \b1^T\exp(\bolds\omega),
\end{equation}
where $\bolds\omega= \log E(\b Y) = \log\bolds\mu$  and
$
\b C_T \log(\b M_T \bolds\mu) = \b{Z}_r\bolds\beta_r
$, from \eqref{eq:redmod}.
The maximization of this likelihood under the above linear constraints
has been discussed by several authors; see
\cite{glomcc1995,lang1996,berrud2002,barcolfor2007},
among others.
\begin{exa}
We give a simple illustration based on an application to data from the
US General Social Survey \cite{gss2007}, for years 1972--2006. The
data are
collected on 13\,067 individuals on 5 variables. There are
three binary responses
concerning individual opinions (1${}={}$favor, 2${}={}$oppose) on legal
abortion if pregnant as a result of rape, $A$; on
death penalty for those convicted of murder, $C$; and on the
introduction of police permits for buying guns, $G$. The potentially
explanatory variables considered are $J$, job satisfaction
(with three levels: 1${}={}$very satisfied, 2${}={}$moderately satisfied, 3${}={}$a
little or very
dissatisfied), and $S$, gender (1${}={}$male, 2${}={}$female). We can interpret
responses $G$ and $C$
as indicators of the attitude towards individual safety, while $C$ and
$A$ are indicators
of the concern for the value of human life, even in extreme situations.

\begin{table*}
\tabcolsep=0pt
\centering
\caption{Multivariate regression chain graph model selection for GSS data.
Model  {(1)} is the pure independence model of Figure~\protect\ref{fig:3}
for $p_{GCA|JS}$.
Models   (2)--(7)  are fitted during the suggested model selection procedure.
On the right are the fitted parameters for the best selected model}
\label{tab:2}
\begin{tabular*}{\textwidth}{@{\extracolsep{\fill}}lll@{\hspace{6pt}}lllll@{}}
\hline
Model for $p_{GCA|JS}$ & Deviance & d.f. & Logit & Const. &
$J_{\operatorname{mdr}}$ & $J_{\mathrm{full}}$ & $S_{\mathrm{f}}$ \\ \hline
(1) $G \ci A | J, S$ and $G \ci J | S$ &12.84 & 10 & $G$ & \hphantom{$-$}0.766 & & &
\hphantom{$-$}0.766 \\
(2) No 5-factor interaction & \hphantom{3}0.49 & \hphantom{1}2 & $C$ & \hphantom{$-$}$1.051$ & \hphantom{$-$}$0.150$ &
\hphantom{$-$}$0.257$ & $-0.458$ \\
(3) $+$ no 4-factor interactions & \hphantom{3}5.59 & 11 & $A$ & \hphantom{$-$}$1.826$ & $-0.033$
& $-0.245$ & $-0.172$ \\
(4) $+$ no 3-factor interactions & 30.16 & 27 & $GC$ & $-0.303$ & & & \\
(6) $+$ Delete edge $GA$ & 33.38 & 28 & $CA$ & \hphantom{$-$}$0.557$ & & & \\
(7) $+$ Delete edge $GJ$ & 34.25 & 30 & & & & & \\
\hline
\end{tabular*}
\end{table*}

The two explanatory variables
turned out to be independent
(with a likelihood ratio test statistic of $w = 0.79$, 1 d.f.). Hence,
we concentrate on the model for the
conditional distribution $p_{GCA|JS}$. Here the saturated model \eqref{eq:modT}
has 42 parameters and the structure of the parameters is that of
Table~\ref{tab:1},
with the only modification of the
dimensions of the interaction parameters involving the factor $J$, with
three levels.
We describe a hierarchical backward selection strategy. For this,
we examine first the sequence of models obtained by
successively removing the higher order interactions;
see Table~\ref{tab:1}. Then we
drop some of the remaining terms to fit independencies.

The results are shown in Table~\ref{tab:2}. The model
with no interactions of an order higher than three has
a deviance of $30.16$ with 27 degrees of freedom adequate.
From the edge exclusion deviances, we verify that we can remove the
edges $GA$ ($w = 33.38-30.16 = 3.22$, 1 d.f.)
and $GJ$ ($w = 34.25-33.38 = 0.87$, 2 d.f.).
The final multivariate regression chain graph model,
as shown in Figure~\ref{fig:3}(a), has a combined deviance of
$34.25 + 0.79 = 35.04$ on 32 degrees of freedom.

%
\begin{figure}[b]

\includegraphics{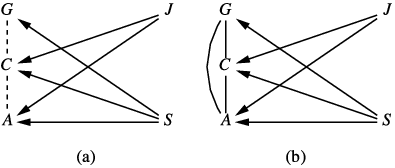}

\caption{{(a)} The multivariate regression chain graph model fitted to
GSS data (Deviance${}={}$13.63, d.f.${}={}$12). The final fitted model
including further non-independence constraints
has a Deviance${}={}$35.04 on $32$ d.f.  {(b)} the best fitting LWF chain
graph model (Deviance${}={}$12.81, d.f.${}={}$18).}
\label{fig:3}
\end{figure}

Notice that the model includes independence and non-independence
constraints, the latter following our preference for a model
with all interpretable parameters. The chain graph model corresponding
exactly to the implied independencies has far more parameters, with
a deviance of $12.84 + 0.79 = 13.63$ against 12 degrees of freedom.
While this model is adequate, the chosen model has a
simpler interpretation. The fitted parameters are shown in Table~\ref
{tab:2} on the
right. The first three rows give the parameters of three univariate
logit regressions
for being in favor of the issue. $J_{\operatorname{mdr}}, J_{\mathrm{
full}}$ measure the effect of moderate
and full job satisfaction, respectively, with respect to a baseline
level of no satisfaction, and
$S_{\mathrm{ f}}$ is the effect of females.
Thus the effect of increased job
satisfaction whatever the gender, is to increase the probability of
being in favor of capital punishment and against abortion. Women are
more favorable than males toward gun
regulation and are more against the death penalty and abortion, all
things being equal. The negative
residual association between $G$ and $C$ and the positive one between
$C$ and $A$
having accounted for gender and job satisfaction are as expected.
As a comparison, in this example, a best-fitting classical chain graph
model with LWF interpretation has one additional edge, as shown in
Figure~\ref{fig:3}.
The multivariate regression chain graph has a simpler interpretation
in terms of three additive logistic regressions and two residual associations
interpretable as deriving from two latent variables.
\end{exa}


\begin{appendix}\label{appm}
\renewcommand{\theequation}{\arabic{equation}}

\section*{Appendix: Proofs}
We shall assume for the joint distribution the existence of a density
with respect
to a~product measure. Proofs using only basic properties
of conditional independence can also be given, but are omitted for brevity.
%
\begin{lemma}\label{prop:iv}
The block-recursive Markov property of type IV is equivalent to
the following three statements: for
all $T \in\tcal$
\begin{subequations}
%
\begin{eqnarray}\label{eq:iv0}
p_{T|\pre(T)} &=& p_{T|\pa_D(T)},  \\\label{eq:iv1}
 p_{A|\pa_D(T)} &=& p_{A|\pa_G(A)}  \qquad \mbox{for all connected } A
\subseteq T,
\\\label{eq:iv2}
 p_{A|\pa_D(T)} &=& \prodx_j p_{A_j|\pa_D(T)} \qquad \mbox{for all disconnected }
A \subseteq T,
\end{eqnarray}
\end{subequations}
\noindent
where $A_j$, $j = 1, \ldots , r$, are the connected components of $A$, if
disconnected.
\end{lemma}

%
%
\begin{pf}
Condition \textsc{(iv0)} states that the joint probability
distribution obeys the \textit{local directed Markov property} relative
to the directed
graph $D$ of the chain components. Then, using the equivalence
of the local and well-ordered local Markov property in directed graphs
applied to the graph of the components as
discussed in \citep{drtplm2008}, Appendix A, \textsc{(iv0)} turns out
to be equivalent to \eqref{eq:iv0}
for any ordering of the components consistent with the chain graph.
Moreover, condition \textsc{(iv2)} has been proved by \cite
{drtric2008} to be equivalent to the joint independence \eqref{eq:iv2}.
Statement \textsc{(iv1)} implies \eqref{eq:iv1} but it can be
restricted to connected subsets $A$ because,
for disconnected subsets, it follows from \eqref{eq:iv2} and from \eqref
{eq:iv1} restricted
to connected sets. If $A$ is disconnected, \eqref{eq:iv2} implies
%
\begin{equation}
p_{A|\pa_D(T)} = \prodx_j p_{A_j|\pa_D(T)} = \prodx_j p_{A_j|\pa
_G(A_j)} = \prodx_j p_{A_j|\pa_G(A)}
\end{equation}
by applying \eqref{eq:iv1} to the connected sets $A_j$ and noting that
$\pa_G(A_j)\subseteq\pa_G(A)$.
Therefore, $p_{A|\pa_D(T)} = p_{A|\pa_G(A)}$ and equation \textsc
{(iv1)} follows.
\end{pf}

Then we are ready to prove that the multivariate regression Markov
property is
equivalent to the above block-recursive Markov property.


\begin{pf*}{Proof of Theorem~\ref{theo:1}}
We establish the equivalence of \eqref{eq:mr1} and \eqref{eq:mr2} with
\eqref{eq:iv0}, \eqref{eq:iv1} and \eqref{eq:iv2} of Lemma~\ref{prop:iv}.

(Definition~\ref{def:1} implies Definition~\ref{def:drton}.)
Equation \eqref{eq:mr1} implies $p_{A|\pre(T)} = p_{A|\pa_D(T)}$ for
all connected $A$
because $\pa_G(A) \subseteq\pa_D(T)$. Thus
\eqref{eq:mr1} implies \eqref{eq:iv1} and
\eqref{eq:iv0} for $A = T$, because any $G_T$ is connected, by
definition. Thus,
if $A$ is disconnected,
\eqref{eq:mr2} gives
\[
p_{A|\pre(T)} = \prodx_j p_{A_j | \pre(T)} = \prodx_j p_{A_j | \pa
_D(T)} = p_{A|\pa_D(A)}
\]
and \eqref{eq:iv2} follows.

(Definition~\ref{def:drton} implies Definition~\ref{def:1}.) Statement
\eqref{eq:iv0} implies, for $A\subseteq T$, that $p_{A|\pre(T)} =
p_{A|\pa_D(T)}$.
Thus for all connected $A$, \eqref{eq:iv1} implies $p_{A|\pre(T)} =
p_{A|\pa_G(A)}$,
i.e., \eqref{eq:mr1}. Moreover, if $A\subseteq T$ is disconnected,
\eqref{eq:iv2} implies
\[
p_{A|\pre(T)} = p_{A|\pa_D(T)} = \prodx_j p_{A_j|\pa_D(T)}
= \prodx_j p_{A_j|\pre_D(T)},
\]
that is, \eqref{eq:mr2}.
\end{pf*}

Given a subvector $\b X_M$ of the given random vector $\b X$,
the log-linear expansion of its marginal probability distribution $p_M$
is
%
\begin{equation}
\log p_M(\b i_M) = \sum_{s \subseteq M} \lambda^{M}_s(\b i_s), \label
{eq:loglin0}
\end{equation}
where $\lambda^M_{s}(\b i_s)$ defines the `interaction' parameters
of order $|s|$ in the baseline
parametrization, that is, with the implicit constraint that the function
returns zero whenever at least one of the indices in $\b i_s$ takes the
first level.
\begin{lemma}\label{prop:loglin}
If $\eta^{(A)}(\b i^\star_A| \b i_{\pre(T)})$ is the multivariate
logistic contrast of the conditional
probability distribution $p_{A|\pre(T)}$ for $A$ subset of $T$, then, with
$B = \pre(T)$,
%
\begin{equation}
\eta^{(A)}(\b i^\star_A| \b i_{B}) =
\sum_{b \subseteq B} \lambda^{AB}_{Ab}(\b i_A^\star, \b i_b),
\end{equation}
where $\lambda^{AB}_{Ab}(\b i_A^\star, \b i_b)$ are the log-linear interaction
parameters of order $A \cup b$ in the marginal probability distribution
$p_{AB}$.
\end{lemma}

\begin{pf}
First note that the multivariate logistic contrasts $\eta^{(A|B)}(\b
i^\star_A| \b i_B)$ can be written
%
\begin{equation}
\eta^{(A|B)}(\b i_A^\star|\b i_B) =
\sum_{s \subseteq A} (-1)^{|A\setdiff s|} \log p_{AB}(\b i_s^\star, \b
i_B, \b1_{A \setdiff s}).
\label{eq:etac2}
\end{equation}
Then we express the logarithm of the joint probabilities $p_{AB}$ as
the sum of log-linear interactions
using \eqref{eq:loglin0},
\[
\log p_{AB}(\b i_s^\star, \b i_B, \b1_{A \setminus s}) = \sum_{a
\subseteq A} \sum_{b \subseteq B} \lambda^{AB}_{ab}(\b i_{a \cap
s}^\star, \b1_{a \setminus s}, \b i_b)
= \sum_{a \subseteq s} \sum_{b \subseteq B} \lambda^{AB}_{ab}(\b
i_{a}^\star, \b i_b).
\]
Therefore, by substitution into equation \eqref{eq:etac2} we get
\begin{eqnarray*}
\eta^{A|B}(\b i_A^\star| \b i_B) &=& \sum_{s \subseteq A}
(-1)^{|A\setdiff s|}
\sum_{a \subseteq s} \sum_{b \subseteq B} \lambda^{AB}_{ab}(\b
i_{a}^\star, \b i_b) \\
&=& \sum_{b \subseteq B} \sum_{s \subseteq A} (-1)^{|A\setdiff s|}
\sum_{a \subseteq s} \lambda^{AB}_{ab}(\b i_{a}^\star, \b i_b)
=\sum_{b \subseteq B} \lambda^{AB}_{Ab}(\b i_A^\star, \b i_b),
\end{eqnarray*}
where the last equality is obtained using a M\"obius inversion; see
\citep{lau1996}, Lemma A.2, page 239.
\mbox{\quad}
\end{pf}

Lemma~\ref{prop:loglin} is used in the proof of Theorem~\ref{prop:base}
given below.
\begin{pf*}{Proof of Theorem~\ref{prop:base}}
If \eqref{eq:constr1} holds for any chain component $T$, then for any
connected set $A \subseteq T$,
$\bolds\eta^{(A)}(\b i_{\pre(T)})$ is a function
of $\b i_{\pa_G(T)}$ only. Therefore, using the diffeomorphism
and the property of upward compatibility discussed in Remark~\ref{rem:diff},
the conditional distribution
$p_{A|\pre(T)}$ coincides with $p_{A|\pa_G(A)}$ and condition \textsc{(mr1)}
holds.

Conversely, if condition \textsc{(mr1)}
holds and $p_{A|\pre(T)} = p_{A|\pa_G(A)}$, for all connected subsets
$A$ of $T$,
then the components of $\bolds\eta^{(A)}(\b i_{\pre(T)})$ are
\begin{eqnarray*}
\eta^{(A)}\bigl(\b i_A^\star|\b i_{\pre(T)}\bigr) &=& \sum_{s \subseteq A}
(-1)^{|A\setdiff s|}
\log p\bigl(\b i_s^\star, \b1_{A \setdiff s}\mid\b i_{\pa_G(T)}\bigr)\\
&=& \sum_{b \subseteq B} \lambda^{AB}_{Ab}(\b i_A^\star, \b i_b), \qquad  \mbox
{with } B = \pa_G(T)
\end{eqnarray*}
by Lemma~\ref{prop:loglin}, and thus \eqref{eq:constr1} holds with
$\bolds\beta^{(A)}_b(\b i_b) = \bolds\lambda^{AB}_{Ab}(\b i_b),$ where
$\bolds\lambda^{AB}_{Ab}(\b i_b)$ denotes the vector
of log-linear parameters $\lambda^{AB}_{Ab}(\b i^\star_A, \b i_b)$ for all
$\b i^\star_A \in\ical^\star_A$.

Condition (\textsc{mr2}) of Definition~\ref{def:1} is equivalent
to imposing that, for any chain component~$T$, the conditional
distribution $p_{T|\pre(T)}$ satisfies the independence model of a~%
covariance subgraph $G_T$. In \cite{kau1997} and \cite
{lupmarber2009} it is proved that,
given a~joint distribution $p_T$, a~covariance
graph model is satisfied if and only if, in the multivariate
logistic parameterization $\bolds\eta_T$, $\bolds\eta^{(A)}= \b0$ for
all disconnected sets $A \subseteq T$. Therefore, extending this
result to the conditional distribution $p_{T|\pre(T)}$ and
considering the diffeomorphism \eqref{eq:satmo}, condition
(\textsc{mr2}) holds if and only if $\bolds\eta^{(A)}(\b i_B) = \b0$
for every disconnected set $A \subseteq T$. Following the
factorial model (7), $\bolds\beta^{(A)}_b(\b i_b) = \b0$ with $b
\subseteq\pre(T)$ for each disconnected subset $A$ of $T$.
Notice that, by Lemma~\ref{prop:loglin}, $\bolds\beta^{(A)}_b(\b i_b) =
\bolds\lambda_{Ab}^{AB}(\b i_b)=\b0$, with $b \subseteq\pre(T)$.
\end{pf*}
\end{appendix}

\section*{Acknowledgements}
We thank Nanny Wermuth, D. R. Cox and two referees for helpful comments.
This work was supported by MIUR, Rome, under the project PRIN 2007,
XECZ7L001/2.

\printhistory

\end{document}